\RequirePackage{fix-cm}
\RequirePackage{rotating}
\documentclass[figuresright,smallextended]{svjour3}      
\smartqed  
\usepackage[numbers]{natbib}
\usepackage[colorlinks=true,urlcolor=blue,citecolor=blue]{hyperref}
\usepackage{doi}
\usepackage{graphicx}
\usepackage{pdflscape}
\usepackage{rotating}
\usepackage{caption}
\usepackage[misc,geometry]{ifsym} 
\usepackage{longtable}
\usepackage{pdflscape}
\usepackage{float}
\usepackage{footnote}
\usepackage[usenames,dvipsnames]{color}

\usepackage[utf8]{inputenc}
\usepackage{graphicx}
\usepackage{multirow}
\usepackage{lineno,hyperref}
\usepackage{booktabs}
\usepackage{array, caption, threeparttable}
\usepackage{etoolbox}
\newcommand*{\affaddr}[1]{#1} 
\newcommand*{\affmark}[1][*]{\textsuperscript{#1}}
\usepackage{float}
\begin{document}
\pagewiselinenumbers
\switchlinenumbers
\nolinenumbers
\title{A Survey on Extraction of Causal Relations from Natural Language Text}

\author{%
\makebox[\linewidth][l]{
Jie Yang \affmark[] \and Soyeon Caren Han\affmark[] \and Josiah Poon\affmark[]
}}
\authorrunning{Jie Yang\and Soyeon Caren Han\and Josiah Poon}

\institute{
            Jie Yang \\
            {jyan4704@uni.sydney.edu.au}\\
            \\
           \Letter Soyeon Caren Han \\
            {caren.han@sydney.edu.au} \\
            \\
            Josiah Poon\\
            {josiah.poon@sydney.edu.au} \\
            \\
            \affaddr{\affmark[]School of Computer Science, The University of Sydney \\
            1 Cleveland Street, NSW 2006, Australia}
}
\date{Received: date / Accepted: date}
\maketitle

\begin{abstract}
As an essential component of human cognition, cause-effect relations appear frequently in text, and curating cause-effect relations from text helps in building causal networks for predictive tasks. Existing causality extraction techniques include knowledge-based, statistical machine learning (ML)-based, and deep learning-based approaches. Each method has its advantages and weaknesses. For example, knowledge-based methods are understandable but require extensive manual domain knowledge and have poor cross-domain applicability. Statistical machine learning methods are more automated because of natural language processing (NLP) toolkits. However, feature engineering is labor-intensive, and toolkits may lead to error propagation. In the past few years, deep learning techniques attract substantial attention from NLP researchers because of its' powerful representation learning ability and the rapid increase in computational resources. Their limitations include high computational costs and a lack of adequate annotated training data. In this paper, we conduct a comprehensive survey of causality extraction. We initially introduce primary forms existing in the causality extraction: explicit intra-sentential causality, implicit causality, and inter-sentential causality. Next, we list benchmark datasets and modeling assessment methods for causal relation extraction. Then, we present a structured overview of the three techniques with their representative systems. Lastly, we highlight existing open challenges with their potential directions.
\end{abstract}

\section{Introduction}
With the rapid growth of unstructured texts online, information extraction (IE) plays a vital role in NLP research. It automatically transforms and stores unstructured texts into machine readable data \cite{cowie1996information}. The complex syntax and semantics of natural language and its extensive vocabulary make IE a challenging task. IE is an aggregation of tasks, which includes named entity recognition (NER), relation extraction (RE), and event extraction. RE refers to extracted and classified semantic relationships, such as whole-part, product-producer, and cause-effect from text. Specifically, the cause-effect relation, which refers to a relationship between two entities \textit{e1} and \textit{e2}, that the occurrence of \textit{e1} results in the occurrence of \textit{e2}, is essential in many areas. For example in medicine, the decision to provide a treatment is based on the relationship that the treatment leads to an improvement in patient’s condition. Or, the critical issues of whether a disease is the reason for a symptom depend on if there are cause-effect relation between them. Extracting such kinds of causal relations from medical literature can support constructing a knowledge graph, which can assisting doctors in quickly finding causality, like \textit{diseases-cause-symptoms, diseases-bring-complications, treatments-improve-conditions}, and finally customize treatment plans. Similarly, extracting cause-effect relations from text, which is the study of causality extraction (CE), has received ongoing attention in media \cite{balashankaretal2019identifying,chang2006incremental,Cole2006A,KHOO1998,Edoardo2017Event}, biomedical \cite{BuiExtracting,khoo2000extracting,mihaila2014semisupervised,sun2017chemicalinduced}, emergency management \cite{qiu2017extracting}, etc.

\begin{figure}[H]
\centering
\includegraphics[width=10cm,height=5cm]{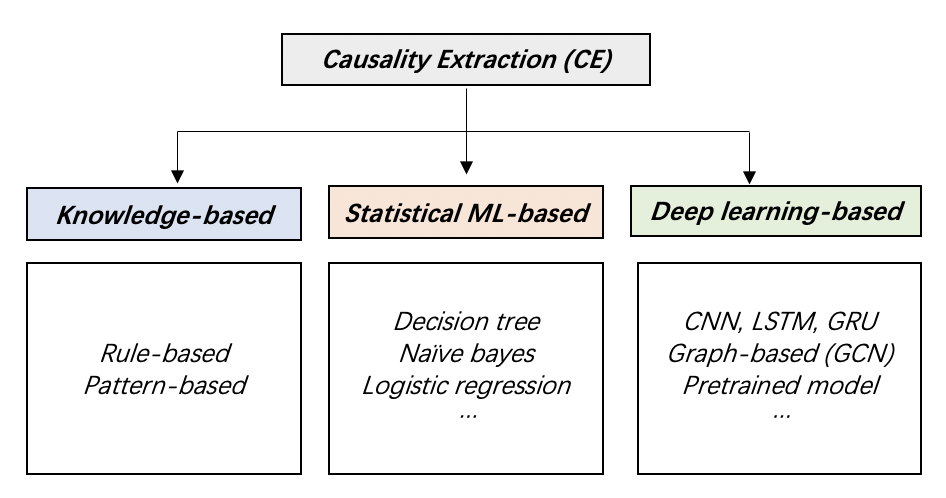}
\caption{Taxonomy of techniques}
\end{figure}

The task of CE focuses on developing systems for identifying cause-effect relations between pairs of labeled nouns from text \cite{beamer2008automatic}. From the aspect of techniques, as shown in Figure 1, there has been a considerable body of CE systems that can be divided into three groups: knowledge-based approaches, statistical machine learning(ML)-based approaches and deep learning-based approaches. Alternatively, CE studies can be classified in terms of different representation patterns: explicit or implicit causality, intra- or inter-sentential causality. Explicit causality has relations that are connected by the following explicit causal connectives: (a) causal links (e.g., \textit{so, hence, therefore, because of, on account of, because, as, since, the result was}); (b) causative verbs (e.g., \textit{break, kill}); (c) resultative phrases; (d) conditional, i.e., \textit{if...then...} and (e) causative adverbs and adjectives \cite{Khoo2002facets}. Implicit causality means explicit causal valence are replaced by ambiguous connectives, e.g., \textit{as, after} in the first four examples, or even without any connectives, as the last example in Table 1. Readers need to use background knowledge to analyzing and reasoning if there is causality in the text. In intra-sentential causality, the “cause” and the “effect” lie in a single sentence, while in inter-sentential causality, the “cause” and the “effect” lie in different sentences. Most CE approaches, like \cite{Garcia1997COATIS,girju2003automatic,khoo2000extracting,kyriakakis2019transfer,mihaila2014semisupervised,radinsky2012learning} identify causality in the basic levels, which are explicit and/or intra-sentential forms. However, causality in many texts is implicit and/or inter-sentential conditions, which are more complicated than basic kinds of causality. Table 2 lists three examples, which include the sentences, causality forms, and the causality pairs.

The rest of the article is structured as follows. We review in detail of previous surveys in Section 2. The benchmark datasets and evaluation metrics for CE system are presented in Section 3 and Section 4, respectively. Then, we survey representative CE systems and summarize them in Section 5 and Section 6. We propose three open problems of the CE task with their potential solutions in Section 7, and the conclusion of this paper are in Section 8. 

\begin{table}[!htbp]
\begin{center}
\setlength{\abovecaptionskip}{0pt}
\setlength{\belowcaptionskip}{0pt}
\caption{Examples for implicity causality.}
\begin{tabular}{p{2cm}p{6cm}p{2.5cm}}
\hline \bf Connectives &\bf Sentences & \bf Labels\\\hline
as & There was no debate \textit{as} the Senate passed the bill on to the House. \cite{Blanco2007Causal}& Causal\\
as & It has a fixed time, \textit{as} collectors well know. \cite{Blanco2007Causal}& Non-causal\\
after & Bischoff in a round table discussion claimed he fired Austin \textit{after} he refused to do a taping in Atlanta. \cite{Mart2017Neural}& Causal\\
after & In stark contrast to his predecessor, five days \textit{after} his election he spoke of his determination to do what he could to bring peace. \cite{Mart2017Neural}& Non-causal\\
- & He derives great joy and happiness from cycling. \cite{beamer2008automatic} & Causal \\
\hline
\end{tabular}
\end{center}
\end{table}

\begin{table}[!htbp]
\begin{center}
\setlength{\abovecaptionskip}{0pt}
\setlength{\belowcaptionskip}{0pt}
\caption{The forms of causal relations.}
\label{tab:aStrangeTable}
\begin{tabular}{p{5.5cm}p{2.5cm}p{2.5cm}}
\hline \bf Sentences & \bf Causality Forms & \bf Causality Pairs\\ \hline
Financial stress is one of the main causes of divorce. &
Explicit with Intra-sentential&  \textit{$<$Financial stress,divorce$>$}\\
Financial stress can speed divorce up.& Implicit & \textit{$<$Financial stress,divorce$>$}\\
You may hear that unfaithful can lead to divorce. On the other hand, financial stress is another significant factor.&Inter-sentential& \textit{$<$Financial stress,divorce$>$} \\
\hline
\end{tabular}
\end{center}
\end{table}

\section{Previous surveys}
With limited exceptions, there is a notable paucity of surveys focusing specifically on CE. It may be because cause-effect is a common relation that researchers scales up to RE literature reviews. Examples include the generalized survey \cite{ZhangReview}, detailed analyses of RE in the biomedical domain \cite{Kadir2013Overview,ZhouBiomedical}, and a survey about the application of distant supervision on RE \cite{Smirnova2019Relation}. From our point of view, however, CE is different from RE, as the former task is a binary classification while the later is multiple classification problem. Meanwhile, the two tasks focus on different kinds of linguistic patterns or features. For example, the punctuation feature can be used in RE to indicate the relation of \textit{Description} and \textit{Attribution}, but it is useless in the task of CE \cite{keskes2014learning}. Also, RE faces the challenge of extracting relations on open-domain corpora, that is, the relation types may not be pre-defined \cite{wuetal2019open}, while the target of CE is clear and there are no new relation types.

In 2016, \citet{AsgharAutomatic} separates CE applications into non-statistical techniques, and statistical and machine learning techniques. Besides reviewing previous approaches, another contribution is the analysis of strengths and weaknesses of the two categories. Early non-statistical methods suffer from constructing annotated linguistic and syntactic patterns manually, while ML-based systems can utilize a small set of seed patterns with algorithms to find these language patterns automatically. Also, most non-statistical models restricted their corpora to a particular domain with a specific text type (e.g., narrative, prose, drama). In comparison, the statistical ML techniques provide better generalization to other domains and types of text. Meanwhile, unlike non-statistical architectures that only extracted explicit cause-effect relations, a large number of ML systems (e.g., \cite{bethardmartin2008learning,rink2010,Sorgente2013Causality,Yang2014}) have the capability to explore implicit relations. In the same year with the study of  \cite{AsgharAutomatic}, \citet{BarikEvent} categorize existing approaches into four groups: CE using handcrafted patterns, CE using semi-automatic causal patterns, CE using supervised learning, and statistical methods. From their point of view, instead of using manually linguistic clues and domain knowledge, semi-automatic learning acquires lexico-syntactic patterns from a larger corpus automatically. Then, these patterns are used to identify in-domain causal relations or evaluate causal patterns in a semi-automated way. For the supervised learning, there are a large number of corpora that required labeled prior to modeling. The above two surveys provide comprehensive reviews of CE, one of their limitations is the lack of review about recent developments in the field, especially deep learning. Luckily, we will review both of the traditional and modern methods in Section 5.

\section{Benchmark datasets}
As we all know that data is the foundation of experiment. There is a number of datasets which have been previously used for evaluating CE models. In this section, we describe four datasets from general domain and two datasets from biomedical domain, and summarize them in terms of their causality sizes, sources, available condition, balanced condition (X and - represent balanced and imbalanced, respectively), and related works in Table 3.

\begin{itemize}
\item \textbf{SemEval-2007 Task 4:}
It is part of SemEval (Semantic Evaluation), the 4$^{th}$ edition of the semantic evaluation event \cite{SemEval2007}. This task provides a dataset for classifying semantic relations between two nominals. Within the set of seven relations, the organizers split the \textit{Cause-Effect} examples into 140 training with 52.0\% positive data, and 80 test with 51.0\% positive data. This dataset has the following advantages: (a) Strong reputation. SemEval is one of the most influential, largest-scale natural language semantic evaluation competition. As of 2020, SemEval has been successfully held for fourteen sessions, and has a high impact in both industry and academia. (b) Easily accessible. Each relation example with the annotated results is collected in a separate TXT file, which can also reduce the workload of data pre-processing. On the contrary, the main limitation is the small data amount, that 140 training and 80 test examples are far from meeting the needs for developing a CE system.

\item \textbf{SemEval-2010 Task 8:}
Unlike its predecessor, SemEval-2007 Task 4, that has an independent binary-labeled dataset for each kind of relation, this is a multi-classification task in which relation label for each sample is one of nine kinds of relations \cite{hendrickx-etal-2010-semeval}. Within the 10,717 annotated examples, there are 1,003 training with 13.0\% positive data, and 328 test with 12.0\% positive data. This small sample amount and imbalanced condition are the major limitations of this dataset.

\item \textbf{PDTB 2.0:}
The second release of the penn discourse treebank (PDTB) dataset from \citet{prasad2007penn} is the largest annotated corpus of discourse relations. It includes 72,135 non-causal and 9,190 causal examples from 2,312 Wall Street Journal (WSJ) articles. In addition, there is a type of implicit relation in the dataset known as AltLex (Alternative lexicalization) corpus, in which causal meanings are not expressed by explicit causal lexical markers. However, the authors store PDTB in a complex way, that researchers need to use tools to convert it into easy-to-operate files.

\item \textbf{TACRED:}
Similar to SemEval, the Text Analysis Conference (TAC) is a series of evaluation workshops about NLP research. The TAC Relation Extraction Dataset (TACRED) contains 106,264 newswire and online text that have been collected from the TAC KBP challenge\footnote{https://www.ldc.upenn.edu/collaborations/current-projects/tac-kbp}, during the year from 2009-2014 \cite{zhang2017positionaware}. The sentences are annotated with person- and organization-oriented related type (e.g., \textit{per:title, org:founded}). The main limitation of TACRED is the small number of examples that there are only 269 \textit{cause\_of\_death} instances available for CE task.
\end{itemize}

The above four corpora are collected from large general-purpose texts, like English Wikipedia and WSJ. At the same time, datasets in specific domains are needed to train and evaluate specific CE systems. Here, we list two causality datasets in the biomedical domain. 
\begin{itemize}
\item \textbf{BioInfer:}
\citet{BioInfer2007} introduce an annotated corpus, BioInfer (Bio Information Extraction Resource), which contains 1,100 sentences with the relations of genes, proteins, and RNA from biomedical publications. There are 2,662 relations in the 1,100 sentences, of these 1,461 (54.9\%) are causal-effect. The original data is collected in detail in the XML form, includes sentence with entity markup.
\item \textbf{ADE:}
The corresponding ADE task aims to extract two entities (drugs and diseases) and relations about drugs with their adverse effects (ADEs) \cite{ADE2012,Li2017}. Dataset in the task is collected from 1,644 PubMed abstracts, in which 6,821 sentences have at least one ADE relation, and 16,695 sentences are annotated as non-ADE sentences. Annotators only label drugs and diseases in the ADE sentences, so some studies, like \cite{Li2017}, only use the 6,821 sentences in the experiments.
\end{itemize}

\begin{table}
\begin{center}
\setlength{\abovecaptionskip}{0pt}
\setlength{\belowcaptionskip}{0pt}
\caption{Benchmark datasets.}
\begin{tabular}{p{1.2cm}p{1.2cm}p{1.2cm}p{1.3cm}p{1.4cm}p{1.2cm}p{1.5cm}}
\hline \bf Datasets & \bf Published Years & \bf Causality Sizes& \bf Sources & \bf Availability & \bf Balanced & \bf Related Works\\ 
\hline
SemEval-2007 Task 4&2007&220&Wikipedia&Publicly available\footnotemark[1]&X & \cite{beamer2008automatic,Classification2009}\\
SemEval-2010 Task 8&2010&1,331&Wikipedia&Publicly available\footnotemark[2]& -& \cite{kyriakakis2019transfer,Li2021,Pakray2014An,Sorgente2013Causality,wangetal2016relation,xuetal2015classifying,zhangetal2018graph,ZhaoEvent}\\
PDTB 2.0&2018&9,190&WSJ&License required\footnotemark[3]& -& \cite{Chen2016Implicit,Hidey2016Identifying,Man2017Multi,lin2009recognizing,Mart2017Neural,Edoardo2017Event,rutherford-xue-2014-discovering}\\
TACRED&2018&269&Newswire, Web&License required\footnotemark[4]&-&\cite{zhang2017positionaware,zhangetal2018graph}\\
BioInfer&2007&1,461&PubMed&Publicly available\footnotemark[5]&X &\cite{Airola2008,Chen2020RelabelTN}\\
ADE&2012&6,821&PubMed&Publicly avaiable\footnotemark[6]&-&\cite{bekoulis-etal-2018-adversarial,BEKOULIS201834,ADE2012,KangADE2014,Li2017,wangADE2020,ZhaoADE2020}\\
\hline
\end{tabular}

\end{center}
\end{table}

\footnotetext[1]{https://sites.google.com/site/semeval2007task4/data}
\footnotetext[2]{https://github.com/sahitya0000/Relation-Classification/tree/master/corpus}
\footnotetext[3]{https://catalog.ldc.upenn.edu/LDC2008T05}
\footnotetext[4]{https://catalog.ldc.upenn.edu/LDC2018T24}
\footnotetext[5]{http://mars.cs.utu.fi/BioInfer/?q=download}
\footnotetext[6]{https://sites.google.com/site/adecorpus/}

\section{Evaluation metrics}
To evaluate the performance of a CE system, the following four metrics are commonly used:
\begin{equation}
Precision = \frac{TP}{(TP + FP)}
\end{equation}
\begin{equation}
Recall = \frac{TP}{(TP + FN)}
\end{equation}
\begin{equation}
F-score = \frac{2*TP}{(2*TP + FN + FP)}
\end{equation}
\begin{equation}
Accuracy = \frac{(TP + TN)}{(TP + FP + TN + FN)}
\end{equation}
As many researchers define their CE systems as relation extraction tasks, that is, to determine whether the annotated causal pair in the input text has causality.
Within their evaluation metrics, TP (true positive) is the number of correctly identified causal pairs. FP (false positive) refers to the number of causal pairs identified as non-causal pairs. TN (true negative) is the number of correctly identified non-causal pairs, FN (false negative) is the number of non-causal pairs that are identified as causal pairs. 

Accuracy and F-score have been (and still are) among the most popular adopted metrics in most classification tasks. However, they may generate overoptimistic, misleading results on imbalanced datasets, as they failed to consider the ratio between positive and negative classes \cite{Chicco2020}. In contrast, Matthews correlation coefficient (MCC) \cite{MATTHEWS1975442} views two classes are equal importance. It is high only when the classifier is doing well in both positive and negative classes: 
\begin{equation}
MCC =\frac{(TP*TN - FP*FN)}{\sqrt{\mathstrut (TP + FP)*(TP + FN)*(TN + FP)*(TN + FN)}}
\end{equation}
The MCC has been used for classifier evaluation over imbalanced datasets, as the publication of \cite{10.3389/fgene.2021.669328,10.3389/fphys.2021.658633,10.3389/fimmu.2018.01783}.

The geometric mean \cite{DBLP:conf/icml/KubatM97}, G-mean, also indicates the balance between performances on both classes. A poor performance in positive examples prediction will lead to a low G-mean value, even if negative instances are correctly classified by the classifier \cite{cdi_springer_books_10_1007_978_3_642_04962_0_53}:
\begin{equation}
G-mean =\sqrt{\mathstrut \frac{TP}{(TP + FN)}*\frac{TN}{(TN + FP)}}
\end{equation}
The effectiveness of G-mean for classifier assessment over imbalanced datasets has been shown in many studies, like \cite{10.1145/1321440.1321461,4433808,4302741}.

The entity labelling metrics also applied to evaluate the models. For example, \citet{KHOO1998} use average precision and recall to judge whether the model can identifying both the boundary of cause and effect. \citet{Tirthankar2018Automatic} compare F-score of labeling "C"(cause), "E"(effect), "CC"(causal connectives) and "N"(None) tags with baseline models. Compared with the relation extraction method, evaluate in a labeling way is more suitable for these systems: both cause and effect have more than one word, and there is no entity mask in the original sentence.

Meanwhile, some approaches evaluate their models based on their topics. The studies of \cite{oh2013whyquestion,Oh2016,oh2017multi-column} aim to recognize causality for finding proper answers in why-QA (question-answering) system. So the authors evaluate their models by precision of the top answer (P@N) and mean average precision (MAP), where P@N measures the number of questions that have correct answers in the top-N passages, and MAP measures the quality of the answer passages ranked by systems. \citet{Kim2013} report causality confidence and topic purity to measure the quality for mining causality topics. For causality confidence, they use the p-value of the Granger causality testing \cite{Granger1969} between two variables. For topic purity, they calculate the entropy of cause word distributions and normalize it to the [0, 100] range.

\section{Causal relation extraction methods}
Many researchers have devoted themselves to the study of causality extraction. In the following subsections, we summarize and classify existing methods for causality extraction, based on the underlying techniques and on the causality forms identified.

\subsection{Knowledge-based approaches}
Knowledge-based CE systems can be divided into pattern-based approaches and rule-based approaches. Some of the former systems express linguistic patterns by means of pre-defined graphical patterns, or keywords (e.g., \textit{thanks to, because, lead to}). On the other hand, patterns can also be explored through sentence structure analyses, like lexico-semantic or syntactic analysis. These structure analysis techniques lead to performance improvements, and additionally are more likely to extract implicit causality relations. As to rule-based approaches, while some systems (similarly to pattern-based approaches) rely on patterns or templates to identify candidate causal instances directly, other systems employ a set of procedures or heuristic algorithms on the syntactic structure of sentences. 

In the following paragraphs we will review how the existing systems, described in the literature, employ knowledge-based techniques to extract causality in different forms.

\textbf{Explicit Intra-sentential Causality} 
\citet{Garcia1997COATIS} and \citet{khoo2000extracting} use patterns to identify explicitly expressed causal relations, within a single sentence. The tool developed by Garcia and colleagues, COATIS, extracts causality from French texts through lexico-syntactic patterns based on 23 explicit causal verbs, like \textit{provoke, disturb, result, lead to}. Due to the special attention given to the syntactic positions of causal verbs and their surrounding noun phrases, COATIS achieves a reasonable precision of 85.2\%. Even though it can only be applied to small fragments of French text, it illustrates how to implement a domain-independent CE system via pre-defined patterns. \citet{khoo2000extracting} introduce an approach to explore causality in medical textual databases. The authors use (medical) domain-specific causal knowledge, such as common causal expressions for depression, schizophrenia, AIDS and heart diseases, as linguistic clues. Even though these clues play a key role in improving performance, their domain speciﬁcity results in a system that can only perform well within the medical domain. \citet{radinsky2012learning} propose a Pundit algorithm to generate causality pairs from news articles. This rule-based approach achieves higher automation than the above two pattern-based systems, thanks to the use of the generalization rule, $<$Pattern, Constraint, Priority$>$. The system obtains 70.4\% precision on titles taken from news articles spreading over a period of 150 years. However, since the rules only cover obvious causality cases, this system achieves a poor recall metric value of 10.0\%. Based on the idea that noun - noun pairs can encode the same semantic relation if they have the same or similar sense collocation,
\citet{beamer2008automatic} propose a WordNet-based learning model to capture noun features from WordNet's IS-A backbone. A different approach is followed by \citet{Classification2009} that instead of using manually constructed resources, they introduce a model that automatically constructs a set of patterns using a pattern cluster algorithm. Even though \cite{beamer2008automatic} outperforms \cite{Classification2009} on SemEval-2007 Task 4 by an F-score of 4.8\%, it should be noted that the former system has poor portability, since it may be unfeasible to use WordNet and additional resources in other applications or corpora.

\textbf{Implicit Causality}
\citet{IttooMinimally} develop a minimally supervised method to identify three pre-defined types of implicit causality in an iterative way. The first type involves resultative verbal patterns, which include verbs like \textit{increase, reduce, kill, become}. The second type involves patterns that make cause and effect inseparable. The third one involves non-verbal patterns, like \textit{rise in} and \textit{due to}. One innovation of this work is the fact that such defined causal patterns are acquired from Wikipedia, as exemplified in Table 4. An experimental study involving 32,545 documents in the Product Development Customer Service (PD-CS) domain achieved an 85.0\% F-score, a result which is comparable to those obtained by state-of-the-art systems. In a study published in 2014, \citet{KangADE2014} focus on the extraction of adverse drug effects from the ADE corpus. This system consists of a concept identification module that recognizes drugs and adverse effects in sentences, and a knowledge-based module for identifying whether a relationship exists between the recognized concepts. A rule-based NLP module, which consists of a number of rules, is combined with a dictionary-based concept recognition and normalization tool, namely Peregrine, to recognize relevant concepts.

\begin{table}[!htbp]
\begin{center}
\caption{Examples of causal patterns from Wikipedia in \citet{IttooMinimally} (2013)}
\begin{tabular}{p{1.5cm}p{9.5cm}}
\hline \bf Causal pattern & \bf Linguistic realization \\ \hline
destroy &"short-circuit in brake wiring \textit{destroyed} the power supply"\\
prevent&"message box \textit{prevented} viewer from starting"\\
exceed&"breaker voltage \textit{exceeded} allowable limit"\\
reduce&"resistor \textit{reduced} voltage output"\\
cause &"gray lines \textit{caused} by magnetic influence"\\
induce &"bad cable extension might have \textit{induced} the motion problem"\\
due to &"replacement of geometry connection cable \textit{due to} wear and tear"\\
mar &"cluttered options \textit{mars} console menu"\\
\hline
\end{tabular}
\end{center}
\end{table}

\textbf{Inter-sentential Causality}
\citet{KHOO1998} propose an approach relying on four kinds of causal links and 2,082 causative verbs to construct a set of verbal linguistic CE patterns. A computer program ﬁnds all parts of the document that match any of the linguistic patterns, so it is able to identify causal relations both within a sentence and between adjacent sentences. It achieves an accuracy of 68.0\% on a set of 1,082 sentences taken from the Wall Street Journal (WSJ) newspaper, with many errors caused by complex sentence structure and the lack of inferencing capability from world knowledge. Verbal-linguistic patterns are used to extract relations between mutations in viral genomes (cause) and HIV drugs (effect) in the system of \citet{BuiExtracting}. An initial text retrieval phase sorts out intra- and inter-sentence candidates if there are $<$mutation, relation, drug$>$ triplets. A subsequent text preprocessing phase simplifies candidate sentences, and manually analyzes the existence of causality based on a list of pre-defined keywords. A final relation extraction phase applies eleven causality rules to form linguistic patterns. This system is used in five hospitals to find resistance data in medical literature. However, due to the large number of noun phrases and technical terms, it can be significantly time-consuming and laborious to simplify sentences, which includes removing parenthetical remarks, replacing \textit{known} terms, grouping mutation and drug names, normalizing sentences, and resolving anaphoras.

\subsection{Statistical machine learning-based approaches}
Statistical machine learning-based approaches require less manual pre-defined patterns than knowledge-based approaches. They usually employ third-party NLP tools (e.g., Spacy \cite{spacy2}, Stanford CoreNLP \cite{manningetal2014stanford}, Stanza \cite{qietal2020stanza}) to generate a set of features for a given collection of labelled data, and subsequently use ML algorithms (e.g., support vector machine (SVM), maximum entropy (ME), naïve bayes (NB), and logistic regression (LG)) to perform the relevant classification. In the following paragraphs we explain in more detail how statistical machine learning techniques are used in CE systems.

\textbf{Explicit Intra-sentential Causality}
Inspired by a previous work which uses lexico-syntactic patterns to infer causation in a semi-automatic way, in 2003, \citet{girju2003automatic} proposes a model to detect causal relations in a QA system. This model focuses on the most frequent explicit intra-sentential causality patterns, \textit{$<$NP1, verb, NP2$>$}, where the verb is a simple causative, and then validates those patterns referring to causation through a set of features based on a decision tree (DT). \citet{Blanco2007Causal} identify causality following the \textit{$<$VerbPhrase, relator, Cause$>$} pattern only, where relator is one of \textit{{because, since, after, as}}. The authors use seven kinds of features with a DT on a popular question classiﬁcation dataset known as TREC \cite{TREC}, and achieve an F-score of 91.3\%. Most errors in this system occur when the relator in the pattern is \textit{as} or \textit{after}. This means that the system tends to only perform well when the instances have clear occurrences of the \textit{because} or \textit{since} keywords. \citet{Pakray2014An,Sorgente2013Causality} and \citet{ZhaoEvent} evaluate their models on the corpus of SemEval-2010 Task 8. The first two methods identify plausible causality instances based on some common sentence-level features, e.g., contextual, constituent parse, and dependency parse features, and then use DT and Bayesian inference, respectively, to discard non-causal instances. Based on the idea that similar causal connectives have similar ways of expressing causality, \citet{ZhaoEvent} introduce a new causal connective feature, which is collected from the similarity of the sentence’s syntactic structure, to divide connectives into diﬀerence classes. A Restricted Hidden Naive Bayes (RHNB) is used to process features and the interactions between causal connectives and lexico-syntactic patterns. The performance of these three models, with F-scores of respectively 85.8\%, 73.9\%, and 85.6\%, demonstrates that the identification of appropriate rules and/or features plays a crucial role in machine learning-based frameworks. \citet{Kim2013} combine a probabilistic topic model with time-series causal analysis in order to mine causal topics. It iteratively reﬁnes topics, increasing the correlation between discovered topics with time-series data. In one experiment, speciﬁc topics that were expected to affect the 2,000 presidential election were mined. Figure 2 shows several important issues (e.g., tax cuts, oil energy, and abortion); such topics are typically also cited in politics-related literature, which shows the eﬃciency of this model. \citet{lin2009recognizing} employ four kinds of features, (production rules, dependency rules, word pairs and contextual) with an ME classifier. An experiment on PDTB 2.0 indicates that the production rules feature contributes the most to the RE task, followed by word pairs, dependency rules, and contextual features. However, as cause-eﬀect is the most predominant relation in the training set, this model tends to label uncertain relations as causality instances. Thus, it gives this relation high recall but very low precision, leading to an F-score of only 51.0\%. \citet{rutherford-xue-2014-discovering} employ Brown cluster \cite{brown-etal-1992-class} pairs to represent relations and employ coreference patterns to identify meaningful relations in PDTB 2.0.

\begin{figure}[H]
\centering
\includegraphics[width=11cm,height=5cm]{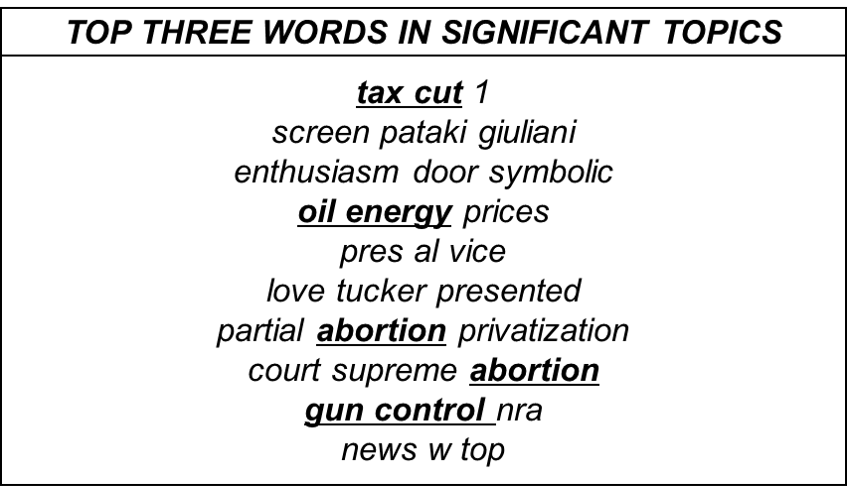}
\caption{Sample results in \citet{Kim2013} (2013)}
\end{figure} 

\textbf{Implicit Causality}
In order to alleviate the shortage of causal connectives, \citet{Hidey2016Identifying} use connectives from PDTB 2.0 AltLex as seed data to identify new alternative lexicalizations from parallel corpora. They train an SVM classifier to process the parallel connectives and lexico-semantic features. The two kinds of features assist the model in achieving the F-score of 75.3\%, which is a significant 11.1\% improvement over its baseline on AltLex corpus. Inspired by the success of kernel-based machine learning methods on RE, \citet{Airola2008} use a dependency-path kernel to extract protein-protein interactions (PPIs) from BioInfer. Each instance is represented by two graphs, one corresponding to the syntactic structure of sentences, and the other to their linear order. In \citet{keskes2014learning}, a ME model is proposed to learn causality in Arabic. Eight linguistic features make signiﬁcant contributions to identifying implicit relations, like the modality feature to check if a sentence has Arabic modal words based on a manually constructed lexicon. The experiment on newswire stories achieves an F-score of 78.1\% and accuracy of 80.6\%. However, these rich and complex feature lists rely heavily on NLP tools like the Standard Arabic Morphological Analyzer and Stanford parser. Unfortunately, due to the specific characteristics of diﬀerent languages, some features that are well extracted may be useless in other languages. \citet{Pechsiri2007} utilize verb-pair rules to train NB and SVM to mine implicit causality from Thai texts. WordNet and pre-deﬁned plant disease information are used to collect the cause and eﬀect verb concepts as a set of verb-pair rules. The experiment on 3,000 agriculture-related sentences obtain precision and recall metrics of 86.0\% and 70.0\%, respectively. Unlike many other methods that use rich sets of features to represent the input instances, this model focuses on the leverage of task-speciﬁc background knowledge, which means that the model can only be applied on a small number of domain-speciﬁc texts. 
in \cite{ADE2012} Gurulingappa (the ADE corpus creator) and colleagues train a ME model with simple features, like words in the sentence, to signal the availability of the corpus. Their experimental results, with an F-score of 70.0\%, are used as baseline performance values for other systems.

\textbf{Inter-sentential Causality}
\citet{marcuechihabi2002unsupervised} utilize lexical pair probability to discriminate causality in inter-sentential forms. They use sentence connecting keywords \textit{Because of} and \textit{Thus} to find candidate sentence pairs, and use pre-collected explicit causality nouns, verbs and adverbs to explore causal lexical pairs. Non-causal lexical pairs are obtained from randomly selected sentence pairs. \citet{oh2013whyquestion} propose a system to explore both intra- and inter-sentential causal relations in a Japanese why-QA system. They utilize regular expressions with explicit keywords in order to identify cue phrases for causality. For each identified cue phrase, the system then extracts three sentences as one causality candidate, including the cue phrase with its left and right sentences. In the process of extracting candidate answers, semantic and syntactic features are used to train a conditional random field (CRF) model to generate cause-effect labels for each word. Finally, to understand chemical induced disease (CID) relations from biomedical articles, \citet{sun2017chemicalinduced} propose two ME models to extract CID at both the intra- and inter-sentential levels, respectively. They construct training and test instances at inter-sentence level complying with three heuristic rules: (a) only pairs of entities that are not involved in any intra-sentential instance are considered at the inter-sentence level; (b) the sentence distance between two entities should be less than three; (c) if there are multiple entities in the same instance, keep the entity pairs with the shortest distance. The authors then use an ME classifier with lexical features to extract this relationship from a collection of 1,500 medical articles.

\subsection{Deep learning-based approaches}
Neural Networks (NNs) are basic algorithms for deep learning (DL). Similarly to a human's neural system, an NN is composed of neurons in three kinds of layers: input, hidden, and output. Each neuron receives input from preceding neurons and produces an output for subsequent neurons. When an NN learns multiple levels of representation from multiple hidden layers, it is said to be a 'deep' neural network, and the process is referred to as 'deep learning' \cite{OAn}.

Compared with knowledge-based and statistical ML-based models, deep learning models map words and features into low-dimensional dense vectors, which may alleviate the feature sparsity problem. Furthermore, the use of an attention mechanism to selectively concentrate on relevant aspects, while ignoring others, tends to make deep learning models more effective. The most typical deep learning models include convolutional neural networks (CNNs), recurrent neural networks (RNNs), and variants of the latter like long short-term memory (LSTM) and gated recurrent units (GRU). Later, the introduction of unsupervised pretraining language models (PTMs) like BERT \cite{Jacob2018BERT}, which return contextualized embeddings for each token, signiﬁcantly improved the performance on many NLP tasks \cite{beltagy2019scibert}. Both CNNs and RNNs can be viewed as sequential-based models, which embed semantic and syntactic information in local consecutive word sequences \cite{yao2019graph}. In comparison, graph-based models, like graph convolutional networks (GCNs) and graph attention networks (GATs), which model a set of points (nodes) and their relationships (edges), have also received the attention of researchers.

In the following paragraphs we discuss how deep learning architectures have been used to solve the CE problem.

\textbf{Explicit Intra-sentential Causality}
The studies of \cite{xuetal2015classifying,Li2021,wangetal2016relation,zhangetal2018graph,kyriakakis2019transfer} employ different deep learning models in order to extract causality from the SemEval-2010 Task 8 dataset. \citet{xuetal2015classifying} use LSTM to learn higher-level semantic and syntactic representations along the shortest dependency path (SDP), while \citet{Li2021} combine BiLSTM with multi-head self-attention (MHSA) to direct attention to long-range dependencies between words. Cases analysis shows that MHSA significantly improves the performance when the causality distance is greater than 10. \citet{wangetal2016relation} propose a multi-level attention-based CNN model to capture entity-specific and relation-specific information. More speciﬁcally, an attention pooling layer is used to capture the most useful convolved features of CNN. The studies of \cite{Li2021,wangetal2016relation} demonstrate the efficiency of attention, especially of the multi-attention mechanism, in the CE task. \citet{zhangetal2018graph} propose a dependency tree-based GCN model to extract relationships. A new tree pruning strategy is applied in order to incorporate relevant information and remove irrelevant context, keeping words that are directly connected to the SDP. Similarly to \cite{xuetal2015classifying}, this technique is based on the definition of the SDP that is used in NLP tools, which will inevitably generate cascading errors. This is the main reason for the F-score of \cite{zhangetal2018graph} to be lower than \cite{wangetal2016relation} by 3.2\%. \citet{kyriakakis2019transfer} explore the application of PTMs like BERT and ELMO \cite{peters-etal-2018-deep} in the context of CE, by using bidirectional GRU with self-attention (BIGRUATT) as the baseline. Experimental results show that PTMs are only helpful for datasets with hundreds of training examples, and that BIGRUATT reaches a performance plateau when thousands of training instances are available. It should be noted that this finding is inconsistent with the results of other studies, which have shown that PTMs are helpful regardless of the magnitudes of the training sets. The TACRED creators, \citet{zhang2017positionaware}, combine LSTM with entity position-aware attention to encode both semantic information and global positions of the entities. The ablation studies show that this position-aware mechanism is effective and pushes the F-score up by 3.9\%. \citet{Edoardo2017Event,Chen2016Implicit,Man2017Multi} focus on causality extraction from the PDTB 2.0 corpus. \citet{Edoardo2017Event} develop a feedforward neural network (FNN) model that combines positional and event-related features with basic lexical features to obtain an enriched feature set. Positional features encode the distance between each word to the \textit{cause} and \textit{effect}, while event-related features account for the semantics of the input sentences. Experimental evaluation of performance indicates that positional features have a positive impact on causality identification. Instead of relying on LSTM only, \citet{Chen2016Implicit} incorporate BiLSTM with GRU in order to capture more complex semantic interactions between text segments. Finally, the method followed in \cite{Man2017Multi} is an attention-based LSTM model that can perform two kinds of representation learning simultaneously: the attention-based module conducts relation representation learning from the interaction between two segments, while a multi-task learning framework uses external corpora to continually improve the performance.

\begin{figure}[H]
\centering
\includegraphics[width=12cm,height=4.5cm]{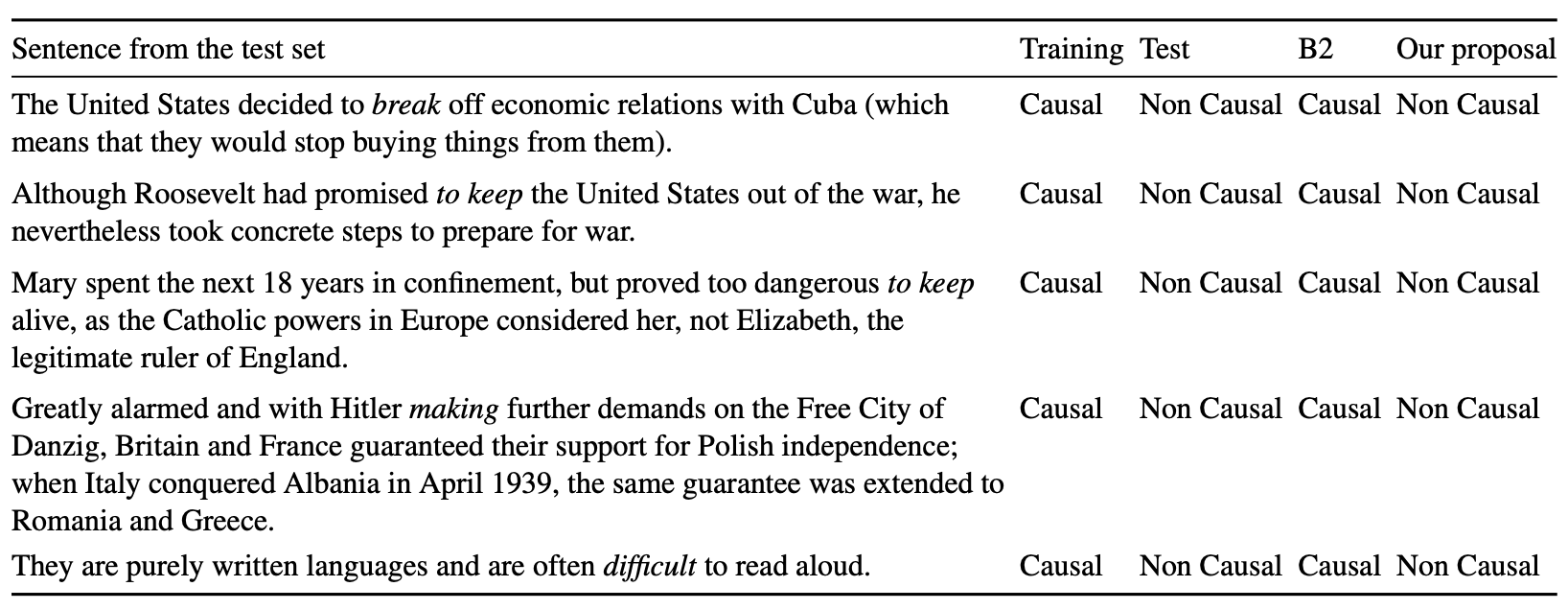}
\caption{Sample results in \citet{Mart2017Neural} (2017)}
\end{figure}

\textbf{Implicit Causality} 
\citet{Mart2017Neural} believe that the use of linguistic features may restrict the ability to represent causality, so they propose an LSTM model incorporating only word embeddings as input. Experimental results obtained on the PDTB 2.0 AltLex corpus show an F-score of 81.9\%. Figure 3 shows ﬁve examples from this study. For instance the ﬁrst example in the figure indicates that the verb \textit{break} mostly has a causal meaning in training examples, but is not a causative verb in the test sentence. It is misclassified by B2 (a baseline model), but correctly classiﬁed by the proposed approach. In the study of \cite{Chen2020RelabelTN}, the authors first incorporate reinforcement learning (RL) to relabel noisy-labeled instances, and then use PCNN, a piece-wise CNN-based model, to iteratively retrain relation extractors with adjusted labels. As the joint entity-relation extraction method can benefit from a close interaction between entities and their relations, \citet{Li2017,wangADE2020,ZhaoADE2020} propose joint models for entity and relation extraction from the ADE corpus. \citet{Li2017} feed character-level representations, word and part-of-speech (POS) embeddings into a BiLSTM to learn entities and context representations. Another BiLSTM is used to learn the relation representation along with the SDP. \citet{wangADE2020} focus on encoding sequence representations and table representations for recognizing entities and their relations, respectively; the two representations then interact with each other attempting to capture task-specific information. The cross-modal attention network (CMAN) in \cite{ZhaoADE2020} is constructed by stacking two attention units, known as BiLSTM-enhanced self-attention (BSA) unit and BiLSTM-enhanced label-attention (BLA) unit, in order to obtain dense correlations over token and label spaces. Another BLA unit captures token-relation interactions to form the ﬁnal label-aware token features. Experimental results on ADE show that CMAN achieves state-of-the-art performance with an F-score of 81.1\%, surpassing \cite{wangADE2020} by 1.0\%. Two other joint models for ADE extraction can be found in \cite{bekoulis-etal-2018-adversarial,BEKOULIS201834}.

\textbf{Inter-sentential Causality}
Taking full advantage of the fact that BiLSTM may alleviate the issue of learning long-range dependencies from a sequence of words, \citet{Jin2020} use CNN to capture essential features from input examples, and then utilize BiLSTM to obtain deeper contextual semantic information between cause and eﬀect. Similarly, after extending the annotation of SemEval2010 Task 8 to phrase-level, \citet{Tirthankar2018Automatic} propose a linguistically informed BiLSTM model to encode word embeddings with linguistic features. The main reason for the misclassfication of causal instances as non-causal instances is that the dependency parser fails to parse the texts correctly, and thus returns improperly syntactic features. \citet{kruengkrai2017improving} introduce a variant of CNN, called multi-column CNN, to recognize event causalities. Based on the assumption that dependency paths between cause and effect can be viewed as background knowledge, they use a wide range of such paths, regardless of whether cause and effect appear within one sentence or in adjacent sentences, taking web texts as extra input. Within this model, diﬀerent columns represent different inputs, such as event causality candidates, contextual information, and background knowledge, with each column having its independent convolutional and pooling layers. All the outputs are concatenated into a SoftMax function to perform the classification. The experimental results demonstrate that related background knowledge significantly improves the performance.

\section{Systems summary}
In the previous section we reviewed 45 systems regarding the different causality forms extracted, i.e., explicit intra-sentential, implicit, and inter-sentential causality, and the different techniques and models employed, i.e. knowledge-based, statistical ML-based, and deep learning-based. Figure 4 contains a statistical summary of the reviewed models. In terms of causality forms, 18.2\%, 31.8\% and 50.0\% of the 45 systems focus respectively on inter-sentential, implicit and explicit intra-sentential causality. In terms of techniques, 20.4\%, 36.4\% and 43.2\% of the systems utilize knowledge-based, statistical ML-based and deep learning-based models. In the next three paragraphs, we separately summarize the advantages and limitations of using each of the three families of techniques.

\textbf{Knowledge-based approaches} These approaches rely on the most straightforward methods, using predefined linguistic rules or patterns to detect whether there exist causal relations hidden in the context. Therefore, they can make use of clear keywords in explicit intra-sentential causality as linguistic clues. On the other hand, since explicit connectives are missing in both implicit and inter-sentential causality, these systems require significant effort to prepare complicated clues, especially involving word-level patterns. Causality extraction by pattern or rule matching can perform well in restricted domains; however, since preparing various kinds of clues is a time-consuming task, knowledge-based methods are unsuitable when the consistency of the dataset is poor.

\begin{figure}[H]
\centering
\includegraphics[width=0.84\textwidth]{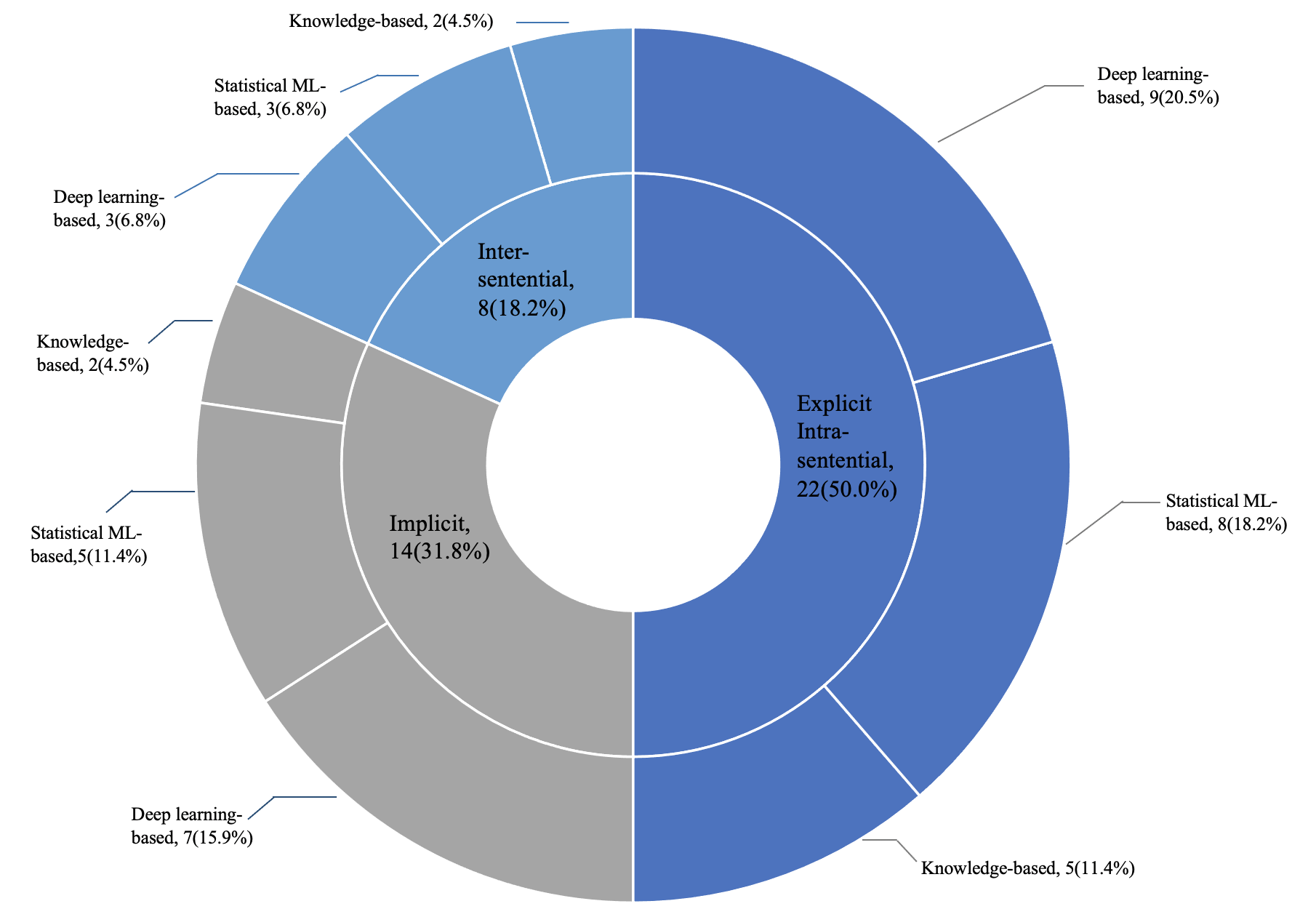}
\caption{Statistical summary of the reviewed systems}
\end{figure}

\textbf{Statistical machine learning-based approaches}
Instead of collecting predefined patterns or rules, traditional ML-based models utilize rich linguistic features or elaborately designed kernels. When the annotated dataset is in the explicit intra-sentential causality form, these systems apply classifiers, with manually or automatically collected features, to remove non-causal instances. When explicit keywords are missing, more complicated features may also need to be prepared. Statistical ML-based approaches usually achieve better performance than knowledge-based systems; however, well-prepared features or kernels may lead to weak portability downstream.

\textbf{Deep learning-based approaches} 
Since deep learning can automatically deduce higher-level information from input vectors and make adjustments to the expected results, these systems are able to focus more on the choice of input features and model architecture, rather than on the preparation of linguistic information. Deep learning-based systems thus have better portability to different applications. However, they require access to larger corpora and more substantial computational resources than the other two techniques.

In conclusion, as illustrated in Figure 5, specific needs and other contextual aspects should be taken into consideration in order to choose an appropriate method for causality extraction.

\begin{figure}[H]
\centering
\includegraphics[width=1\textwidth]{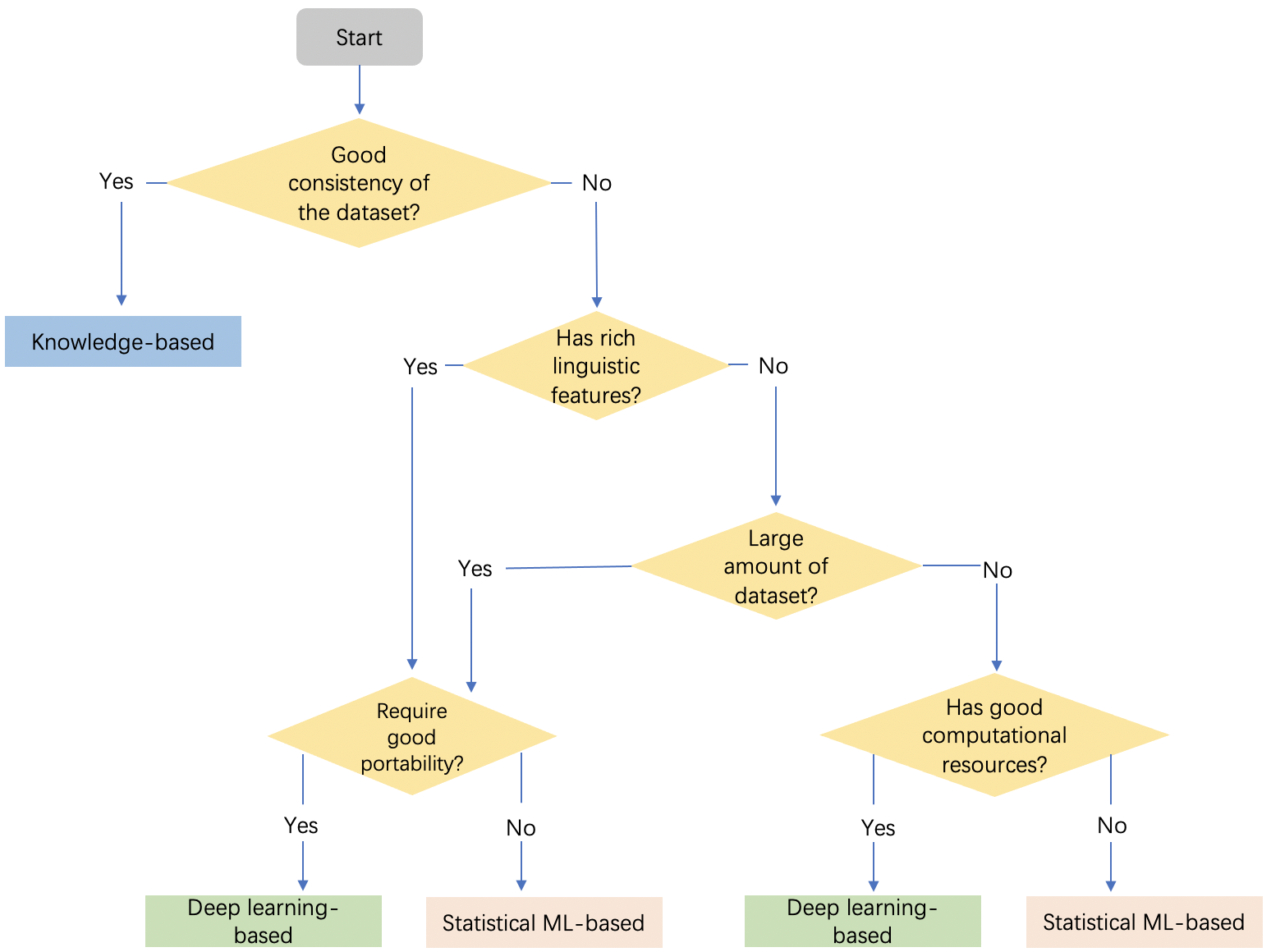}
\caption{The process to choose the appropriate technique for causality extraction}
\end{figure}

Among the reviewed systems, 28 approaches evaluate their models using the benchmark datasets that we introduced in Section 3. From Table 5, we can see that deep learning-based methods achieve new state-of-the-art results, and even show substantial improvements in most of the benchmark datasets. For instance the F-score of \citet{kyriakakis2019transfer} surpass \cite{Sorgente2013Causality} by 16.7\% (on the SemEval-2010 Task 8 corpus), and \cite{ZhaoADE2020} achieves a higher F-score than \cite{KangADE2014} by 25.8\% (on the ADE corpus ). A notable exception is the deep learning-based model of \cite{Chen2020RelabelTN}, which achieves an F-score of only 49.8\% on BioInfer, poorer than \cite{Airola2008} by 11.5\%. The main reason for this is that the model extracts relations after relabeling noise data iteratively, which alerts us to the fact that model performance is closely related to data quality. On the other hand, performance is not the only criterion for judging or choosing an approach. When models based on the same technology have similar or even the same results, we should also take the local situation and needs into account. For example, the F-score of \citet{Li2021} is lower than the F-score of \citet{zhangetal2018graph} by 0.2\% on SemEval-2010 Task 8. However, it is able to avoid the error propagation that is characteristic of the tree pruning strategy, and may have better portability. 

The remaining 17 approaches use their own collected datasets or other kinds of publicly available datasets to evaluate their models’performances. We summarize these approaches in Table 6.

\begin{table*}
\small
\caption{Approaches on benchmark datasets.}
\begin{tabular}{llllll}
\hline
Dataset & System & Year & Technique & F-score(\%)\\
\hline
SemEval-2007&\citet{beamer2008automatic}&2008&Knowledge-based&65.8 \\
Task 4&\citet{Classification2009}&2009&Knowledge-based&\textbf{70.6} \\ \hline
SemEval-2010&\citet{Sorgente2013Causality}&2013&Statistical ML-based&73.9\\
Task 8&\citet{xuetal2015classifying}&2015&Deep learning-based&83.7\\
&\citet{Li2021}&2021&Deep learning-based&84.6\\
&\citet{zhangetal2018graph}&2018&Deep learning-based&84.8\\
&\citet{ZhaoEvent}&2016&Statistical ML-based&85.6\\
&\citet{Pakray2014An}&2014&Statistical ML-based&85.8\\
&\citet{wangetal2016relation}&2016&Deep learning-based&88.0\\
&\citet{kyriakakis2019transfer}&2019&Deep learning-based&\textbf{90.6}\\\hline
PDTB 2.0&\citet{lin2009recognizing}&2009&Statistical ML-based&51.0\\
&\citet{rutherford-xue-2014-discovering}&2014&Statistical ML-based&54.4\\
&\citet{Edoardo2017Event}&2017&Deep learning-based&54.5\\
&\citet{Chen2016Implicit}&2016&Deep learning-based&54.8\\
&\citet{Man2017Multi}&2017&Deep learning-based&\textbf{58.9}\\ \hline
PDTB 2.0&\citet{Hidey2016Identifying}&2016&Statistical ML-based&75.3\\ 
AltLex&\citet{Mart2017Neural}&2017&Deep learning-based&\textbf{81.9}\\ \hline
TACRED&\citet{zhang2017positionaware}&2017&Deep learning-based&65.4\\
&\citet{zhangetal2018graph}&2018&Deep learning-based&\textbf{68.2}\\ \hline
BioInfer&\citet{Chen2020RelabelTN}&2020&Deep learning-based&49.8\\
&\citet{Airola2008}&2008&Statistical ML-based&\textbf{61.3}\\ \hline
ADE&\citet{KangADE2014}&2014&Knowledge-based&54.3\\
&\citet{ADE2012}&2012&Statistical ML-based&70.0\\
&\citet{Li2017}&2017&Deep learning-based&71.4\\
&\citet{BEKOULIS201834}&2018&Deep learning-based&74.6\\
&\citet{bekoulis-etal-2018-adversarial}&2018&Deep learning-based&75.5\\
&\citet{wangADE2020}&2020&Deep learning-based&80.1\\
&\citet{ZhaoADE2020}&2020&Deep learning-based&\textbf{81.1}\\
\hline
\end{tabular}
\end{table*}

\begin{sidewaystable}
\small
\caption{Approaches on other datasets.}
\begin{tabular}{llllll}
\hline
Technique& System & Year & Causality form & Dataset &Performance\\
\hline
Knowledge-based&\citet{KHOO1998}&1998&Inter-sentential&1,082 WSJ sentences&Accuracy (68.0\%)\\
&\citet{khoo2000extracting}&2000&Explicit intra-sentential&130 medical abstracts&Precision (68.0\%)\\
&\citet{Garcia1997COATIS}&2006&Explicit intra-sentential&Technical texts in French&Precision (85.2\%)\\
&\citet{BuiExtracting}&2010&Inter-sentential&630 medical sentences& F-score (84.0\%)\\
&\citet{radinsky2012learning}&2012&Explicit intra-sentential&150 years of news articles &Precision (77.8\%)\\
&\citet{IttooMinimally}&2013&Implicit&32,545 documents in PD-CS&F-score (85.0\%)\\ \hline
Statistical ML-based&\citet{marcuechihabi2002unsupervised}&2002&Inter-sentential&BLIPP&Accuracy (87.3\%)\\
&\citet{girju2003automatic}&2003&Explicit intra-sentential&TREC&Precision (73.9\%)\\
&\citet{Pechsiri2007}&2006&Implicit&3,000 medical sentences in Thai&Precision (86.0\%)\\
&\citet{Blanco2007Causal}&2008&Explicit intra-sentential&TREC&F-score (91.3\%)\\
&\citet{Kim2013}&2013&Explicit intra-sentential&Six months of news articles&t-value (3.87)\\
&\citet{oh2013whyquestion}&2013&Inter-sentential&850 Japanese QA examples&F-score (77.0\%)\\
&\citet{keskes2014learning}&2014&Implicit&90 documents in Arabic&F-score (80.6\%)\\
&\citet{sun2017chemicalinduced}&2016&Inter-sentential&1,500 medical abstracts&Precision (58.3\%)\\ \hline
Deep learning-based&\citet{kruengkrai2017improving} &2017&Inter-sentential&159,350 web sentences&Precision (55.1\%)\\ 
&\citet{Tirthankar2018Automatic} &2018&Inter-sentential&Extended semEval-2010 Task 8&F-score (66.0\%)\\
&\citet{Jin2020} &2020&Inter-sentential&1,986 sentences in Chinese&F-score (82.3\%)\\\hline
\end{tabular}
\end{sidewaystable}

\section{Open problems and future directions}
From the representative systems in Section 5, we can know that causal relation extraction has received increasing attention over the past decade. However, it is a non-trivial problem and many challenges remain unsolved, such as the following three problems: 
\begin{itemize}
\item \textbf{Multiple Causalities:}
Most previous CE only focused on one causal pair from an instance, but causality in the real-world literature is more complex. Causal Patterns in Sciences from Harvard Graduate School of Education introduce three common causal patterns\footnote{https://www.cfa.harvard.edu/smg/Website/UCP/causal/causal\_types.html} as below:
\begin{itemize}
\item (9a) Domino Causality that one cause produces multiple effects.
\item (9b) Relational Causality that two causes work together to produce an effect.
\item (9c) Mutual Causality that cause and effect impact each other simultaneously, or sequentially.
\end{itemize}
Like the study of \cite{Tirthankar2018Automatic}, traditional ways to deal the above kinds of multiple causalities is dividing sentence into several sub-sentences that extracted causal pairs separately. This method is computationally expensive and cannot take into consideration the dependencies among causality pairs. 

The Tag2Triplet algorithm from \cite{Li2021} can extract multiple causal triplets simultaneously. It counts the number and the distribution of each causal tag to judge the tag as simple causality or complexity causality. Afterwards, it applies a Cartisian Product of the causal entities to generate possible causal triplets. In addition, \cite{christopoulou2018,wangetal2019extracting} utilize deep learning with relational reasoning to identify multiple relations in one instance simultaneously.

\item \textbf{Data Deficiency:}
Typically, for many classification tasks, more than 10 million samples are required to train a deep learning model, so that it can match or exceed human performance \cite{Goodfellowetal2016}. However, just as the size of the four benchmark datasets introduced in Section 3 is far from the size of a satisfactory deep learning model, the annotated data in the real-world is very specific and small. 

Based on the assumption that \textit{any sentence that contains a pair of entities that participate in a known Freebase relation is likely to express that relation}, \citet{mintz2009distant} introduce the first distant supervision (DS) system for relation extraction, which creates and labels training instances by Freebase as a relation labels resource. However, this method suffers from a large amount of noise labelled data. The survey of \cite{Smirnova2019Relation} introduces methods of addressing the problem of incomplete and wrong labels from DS, like at-least-one models, topic-based models, and pattern correlations models. The very recent research from \citet{Huang2020} propose a novel way for relation extraction from insufficient labelled data. They first utilize a BiLSTM with attention mechanism to encodes sentences in an unsupervised learning way, the word sequence of entity pairs act as the relation embeddings. Afterwards, a random forest classifier is used to learn the relation types from these relation embeddings. This approach of combine unsupervised learning with supervised learning provide us another new idea of solving data deficiency problem in CE task. 

\item \textbf{Document-level Causality:}
Both intra- and inter-sentential causality are at the sentence-level, in real-world scenarios, however, large amounts of causality span multiple sentences, and even in different paragraphs. Unlike being extracted through linguistic cues or features directly, a satisfactory document-level CE requires that the model has strong pattern recognition, logical and common-sense reasoning \cite{Christopoulou2019ConnectingTD}. All of these aspects need long-term research and exploration. 

\citet{zengEMNLP2020GAIN} introduce a system of combine GCN with relational reasoning to extract relations within a document. They first construct a mention-level GCN to model complex interaction among entities, and then utilize a path reasoning mechanism to infer relations between two entities. This method outperforms the state-of-the-art on the public dataset, DocRED from \citet{yaoetal2019docred}. Similar approaches can be found in \cite{minhtranetal2020dots,wangetal2020global}. 

\end{itemize}
\section{Conclusion}
Causal relations in natural language text play a key role in clinical decision-making, biomedical knowledge discovery, emergency management, news topic references, etc. Therefore, successful causality extraction from fast-growing unstructured text data is a fundamental task towards constructing a causality knowledge base. In this paper, we conducted a comprehensive review of CE in which we introduced six kinds of benchmark datasets and the evaluation metrics for this task. Afterwards, we reviewed existing approaches that use traditional or modern techniques to extract different causality forms. From Section 5 and Section 6, We know that the critical step to extract explicit and implicit causality is to prepare linguistic keywords, patterns, and features. While intra-sentential and inter-sentential causality depend on the way of preparing input instances. Also, we introduced three challenges, which are multiple causalities, data deficiency, and document-level causality extraction, with their potential solutions.

Deep learning provides promising directions for CE tasks. Specifically, domain-related PTMs with graph-based model hold great potential for the two reasons: 1) As the word distributions of general-purpose corpora are quite different with the word distributions of specific domain corpora, the standard PTMs has been shown not to perform well in specialized domains \cite{Jacob2018BERT}. In contrast, pre-training from scratch on domain-specific corpora, like SciBERT \cite{beltagy2019scibert}, BioBERT \cite{lee2019biobert} and BlueBERT \cite{peng2019transfer}, can alleviate this limitation. 2) The study of \cite{guoetal2019attention,zhangetal2018graph} demonstrate the advantage of GCN in complex texts. Thus, we can solve the CE problem by combining domain-specific PTMs with graph models. 

\bibliographystyle{spbasic}   
\bibliography{mybibfile}   
\end{document}